# Modeling of Parallel Single-Pixel Imaging for 3D Reconstruction: New Insights and Opportunities

Feifei Chen, Yunan Shen, Chengmin Liu, Zhaosheng Chen, Xi Tang, Zhengdong Chen, Qican Zhang[*], Zhoujie Wu[*]

College of Electronics and Information Engineering, Sichuan University, Chengdu, Sichuan, China

[*]zqc@scu.edu.cn

[*]zhoujiewu@scu.edu.cn

**Abstract**

The growing prevalence of intelligent manufacturing and autonomous vehicles has intensified the demand for three-dimensional (3D) reconstruction under complex reflection and transmission conditions. Traditional structured light techniques rely on inherent point-to-point triangulation, which limits accurate 3D measurements in these challenging scenarios. Parallel single-pixel imaging (PSI) has demonstrated unprecedented superiority under extreme conditions and has emerged as a promising approach of accurate 3D measurements. However, a complete theoretical model has not been reported in existing work to well explain its underlying mechanisms and quantitatively characterize its performance. In this study, a comprehensive theoretical model for the PSI method is proposed, including imaging and noise models. The proposed imaging model describes light transport coefficients under complex illumination, elucidating the intrinsic mechanisms of successful 3D imaging using PSI. The developed noise model quantitatively analyzes the impact of environmental noise on measurement accuracy, offering a framework to guide the error analysis of a PSI system. Numerical simulations and experimental results validate the proposed models, revealing the generality and robustness of PSI. Finally, potential research directions are highlighted to guide and inspire future investigations. The established theoretical models lay a solid foundation of PSI and brings new insights and opportunities for future application in more demanding 3D reconstruction tasks.

## 1. Introduction

Optical three-dimensional (3D) measurement and sensing plays an important role in numerous fields [1-6]. As one of the representative techniques, structured-light techniques have been sufficiently studied on improving measuring accuracy [7, 8], speed [9-12] and dimensions [13-17] over past several decades.

But all the developed techniques follow point-to-point triangulation, which adopt an assumption that light only travels along one direct path, imposing limitations in more challenging applications under complex illumination [18]. Such kind scene refers to the combination of direct and global illumination components and can be categorized into two types. The first involves scenes with complex reflection conditions, commonly found in industrial inspection and smart assembly. Examples include specular reflections on metal surfaces or mirrors, causing a multipath indirect component, and mixed surfaces (e.g., jade, ice, skin, and wax) result in a subsurface scattering indirect component. The second type includes scenes with complex transmission conditions, often encountered in advanced manufacturing and material sorting. For instance, in metal additive 3D printing, real-time monitoring through dust leads to a volumetric scattering indirect component, while material sorting involves multipath direct illumination when identifying objects behind semi-transparent surfaces. The light propagation paths under complex illumination break the traditional "point-to-point" single reflection condition. Direct utilization of traditional 3D measurement methods may fail due to the aliasing or loss of depth encoding information.

To solve the above-mentioned issue of aliasing between global and direct components, Nayar found that projecting high-frequency structured light patterns can suppress inter-reflection components [19]. And based on this idea, Gupta developed micro phase-shifting methods [20] and XOR Gray code methods to realize 3D measurement under global illumination [21]. Similarly, Moreno et. al. proposed an embedded phase shifting to project single high-frequency fringe to realize absolute measurement [22]. All methods based on high frequencies assume the presence of only low-frequency indirect illumination, which restricts their applicability. Zhang et. al. proposed a bimodal multi-path mathematical model [18] to visualize phase difference as a function of frequency and achieved measurement on step edges and semi-transparent surfaces, but it does not work for subsurface scattering. Light detection and ranging (LiDAR) permits excellent surface-to-surface depth resolution of multi-path scenarios [23-26]. Wallace et. al utilized a scanning sensor to reconstruct the depth profile of an object located behind a wooden trellis fence [27]. Tacchella et. al demonstrated impressive real-time imaging through camouflage nets using a SPAD array and plug-and-play point cloud denoisers [28]. Jiang et. al demonstrated a single-photon array LiDAR system with tailored computational algorithms for 3D imaging through semitransparent materials, achieving video-rate imaging of camouflaged scenes at 20 frames per second (fps) [29]. However, these methods are only able to achieve depth perception or estimation due to the limited accuracy and spatial resolution of time-of-flight method [30, 31]. These mentioned methods still address problem along point-to-point triangulation, making them applicable only to specific tasks. Therefore, a general framework is needed to take it into an overall consideration.

To achieve general measurement under complex illumination conditions, the core task is to describe the light transport process with unpredictable paths. The rapid development of computational optical imaging technology integrates the illumination source, transmission medium, optical system, and imaging detector, offering new perspectives to overcome traditional challenges in imaging [32-34]. As an effective way to describe the equilibrium distribution of radiance in a scene, light transport coefficients (LTC) give the total received radiance at a point in terms of emission from all possible positions on the light source. And it can represent the whole transport process using a four-dimensional transmission matrix [35-37]. O'Toole et al. introduced primal-dual coding to investigate LTC by alternately adjusting programmable masks and illuminations within a single exposure. [38]. Subsequently, an energy-optimized illumination method was achieved through the introduction of a homogeneous encoding technique, enabling 3D reconstruction under complex reflection and transmission conditions, such as specular reflection superposition, strong lighting interference and smoky conditions [39]. This technique effectively captures the intricate transmission process between projector and camera pixels and has been utilized in 3D shape measurement under global illumination. But extra well-designed coaxial arrangement between camera and DMD is essential.

Booming single-pixel imaging (SI) techniques generate images by projecting varying patterns and collecting reflected energy using a single-pixel detector [40-42]. This distinctive working mechanism enables point-to-plane imaging from a computational viewpoint. Based on this technique, Jiang et al. introduced the parallel single-pixel imaging (PSI) method, where each pixel in an array camera is considered as an individual single-pixel detector, and the LTC are reconstructed using the SI algorithm [43]. Then, the global and direct illumination component can be separated and located without any extra hardware cost. Using this technique, mixed scenes with multi-type reflection including diffuse surface, interreflection, subsurface scattering and overexposure surface. However, the need for numerous basis patterns for each SI-based reconstruction limits the practical applications of this technique. To further improve measuring efficiency, Jiang et. al. proposed projective PSI (pPSI) to reduce 4D LTC into multiple projection function based on Fourier slice theorem [44]. 336 patterns are required for one measurement in pPSI, and it achieves measurement under three kinds of global illumination including inter-reflection, subsurface scattering and step edge fringe aliasing. Recently, Wu et. al. proposed multi-scale PSI (MS-PSI) to decrease the projected patterns to 15 [45]. MS-PSI efficiently separates and utilizes different illumination and reflection components, enabling dynamic depth measurement on various surface types, 3D imaging through complex transmission media, and even two-layer 3D imaging [46].

The emerging PSI technique has demonstrated its unprecedented superiority under complex illumination conditions without the need for additional hardware costs. However, current research tends to focus more on showing its measurement capabilities, lacking a systematic imaging model to explain its inner mechanisms and a noise model to quantitatively assess measurement performance and the noise tolerance limit. Thus, a comprehensive theoretical model of PSI is essential to clarify its capabilities and quantitatively evaluate its measurement errors.

In this paper, we establish deterministic imaging and noise model of PSI technique to explain the source of its superiority. The imaging model qualitatively describes the reconstruction mechanism of PSI under complex illumination conditions, providing theoretical support and explanation for its unprecedented measurement capabilities, while also showcasing broader and potential opportunities on more demanding tasks. The noise model quantitatively assesses how environmental noise affects measurement accuracy, providing a structured framework to predict and manage error levels in PSI systems. Numerical simulations and experiments validate the effectiveness of the proposed models and demonstrate the superior performance of PSI under complex illumination conditions. This work lays a strong foundation for PSI, offering new opportunities for its applications in advanced 3D reconstruction tasks, such as implications for intelligent manufacturing and autonomous vehicles.

## 2. Results

### 2.1 Principle of parallel single-pixel imaging

PSI is an emerging technique based on point-to-plane single-pixel localization and its reconstruction process is shown in Fig. 1. The hardware setup consists of a projector and a camera, as illustrated in Fig. 1(a). And a series of Fourier basis patterns are projected to separate aliased complex illumination components. The SI reconstruction algorithm is applied in parallel to each pixel, allowing the illumination component collected from each pixel to be imaged and separated on a two-dimensional plane.

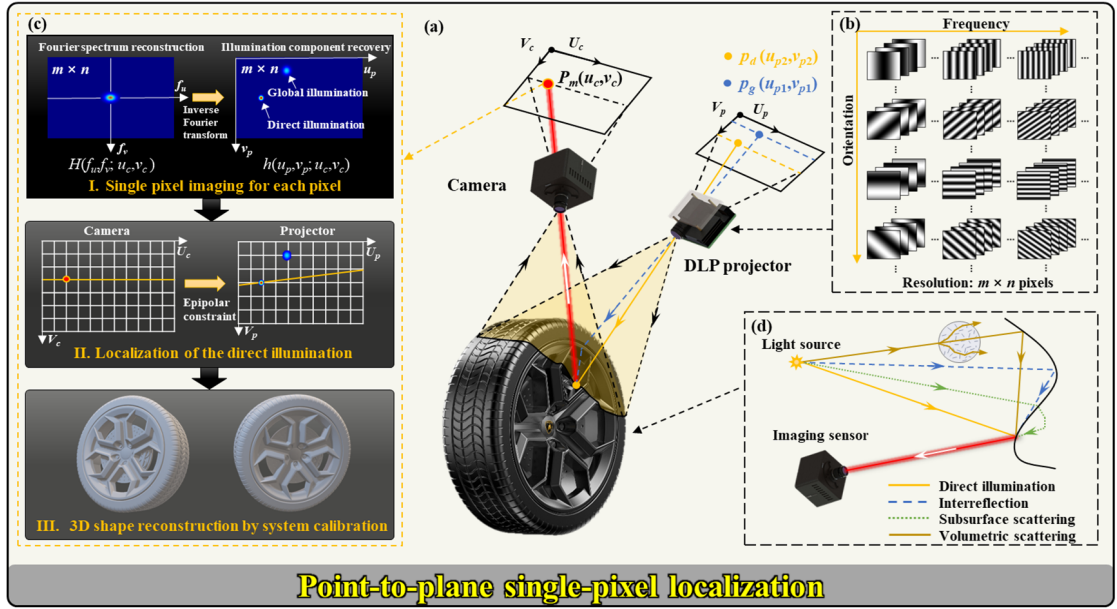

Figure 1. Schematic of PSI for 3D surface reconstruction. (a) Measurement system. (b) Projected sinusoidal patterns. (c) Process of 3D reconstruction using PSI algorithm. (d) Components under complex illumination.

In PSI, the imaging process is modeled using the LTC:

$$I(u_c, v_c) = A(u_c, v_c) + \sum_{v_p=0}^{n-1} \sum_{u_p=0}^{m-1} h(u_p, v_p; u_c, v_c) P(u_p, v_p), \quad (1)$$

where $I(u_c,v_c)$ represents the intensity of the combined illuminations for camera pixel; $A(u_c,v_c)$ represents the environmental illumination; $P(u_p,v_p)$ denotes the intensity for projector pixels. And $h(u_p,v_p;u_c,v_c)$ represents the LTC between camera pixels and projector pixels, describing the intensity transfer ratio between any two pixels in the imaging plane and the projection plane. The LTC for camera pixel $(u_c,v_c)$ can be determined using the Fourier single-pixel imaging (FSI) algorithm. As shown in Fig. 1(b), the projected Fourier basis patterns can be described as follows:

$$P_i(u_p, v_p; f_u, f_v) = a + b\cos[2\pi(f_u u_p + f_v v_p) + \delta_i], i = 1, 2, 3...N, \quad (2)$$

in which, $f_u$ and $f_v$ are the frequency components of the projected patterns in $u$ and $v$ directions, respectively, which can be further calculated as $f_u=k/m$, $k=0,1…m-1$ and $f_v=l/n$, $l=0,1…n-1$. And $a$ is the average, $b$ is the contrast. $N$ represents the phase-shifting step, and $\delta_i$ is given by $2(i-1)/N$. By extending Eq. (1) to Eq. (2), we get:

$$I_i(f_u, f_v; u_c, v_c) = A(u_c, v_c) + \sum_{v_p=0}^{L-1} \sum_{u_p=0}^{K-1} h(u_p, v_p; u_c, v_c) P_i(u_p, v_p; f_u, f_v). \quad (3)$$

The reconstruction of the Fourier spectrum is performed:

$$H(f_u, f_v; u_c, v_c) = \left[\sum_{i=1}^{N} I_i(f_u, f_v; u_c, v_c)\sin\delta_i\right] + \left[\sum_{i=1}^{N} I_i(f_u, f_v; u_c, v_c)\cos\delta_i\right]. \quad (4)$$

After doing fast inverse Fourier transform (noted as $F^{-1}$), The normalized LTC $h(u_p,v_p;u_c,v_c)$ for camera pixel can be determined:

$$h(u_p, v_p; u_c, v_c) = \frac{2}{Nb} \cdot F^{-1}[H(f_u, f_v; u_c, v_c)]. \quad (5)$$

As illustrated in Fig. 1(c), in the projection plane, the direct and global components are distinguished for a specific pixel. According to the triangulation precondition, the direct component will lie on the epipolar plane defined by that pixel [47]. As a result, the direct illumination component's precise position is determined by applying the epipolar constraint and performing subpixel searching. In this manner, a unique correspondence between the imaging and projection planes is established, enabling 3D surface reconstruction through a stereo vision algorithm [48] under complex illumination conditions.

In summary, the unique point-to-plane measurement principle of PSI allows it to effectively separate complex illumination components and eliminate global illumination interference, enabling the solution of more challenging tasks. Unlike traditional methods, the measurement accuracy depends on the positioning precision of direct illumination components rather than phase accuracy. The error analysis framework of this novel architecture offers a potential for improved noise resistance. However, the imaging model and noise model of PSI has not been reported in existing work to well clarify the superiority and quantitively define the noise endurance of PSI. To this end, a comprehensive model of PSI is systematically deduced, including imaging and noise models.

## 2.2 Generalized measurement under global illumination with "point-to-plane" imaging model

After careful consideration, we decided that the imaging model of PSI should be thoroughly established as described in Eq. (6).

$$h(u_p, v_p; u_c, v_c) = \left\{\sum_{k=0}^{M} h^k(u_p, v_p)\delta[u_p - u_p^k, v_p - v_p^k] * R_d^k(u_p^k, v_p^k)\right\} * F^{-1}(1 - M_{mask}). \quad (6)$$

According to the different dominant factors in Eq. (6), three typical illumination conditions can be described as shown in Fig. 2. The LTC $h(u_p,v_p;u_c,v_c)$ can be described as the sum of the multiple illumination components $h^k(u_p,v_p)$, which are modulated by different transmission and reflection conditions. Shifting factor $\delta[u_p-u_p^k, v_p-v_p^k]$ depicts the interreflection occurring on specular surfaces or multi-path direct illumination existing on imaging through semitransparent surface. The diffusion factor $R_d^k(u_p^k,v_p^k)$ describes subsurface scattering reflection condition occurring on translucent materials or biological tissues. And $R_d^k(u_p^k,v_p^k)$ also describes the volumetric scattering transmission condition existing on imaging through scattering media such as smoke and dust. The sampling factor $M_{mask}$ illustrates damaged area in the frequency spectrum of the LTC due to the overexposure condition occurring on highly reflective materials such as metals and specular surfaces.

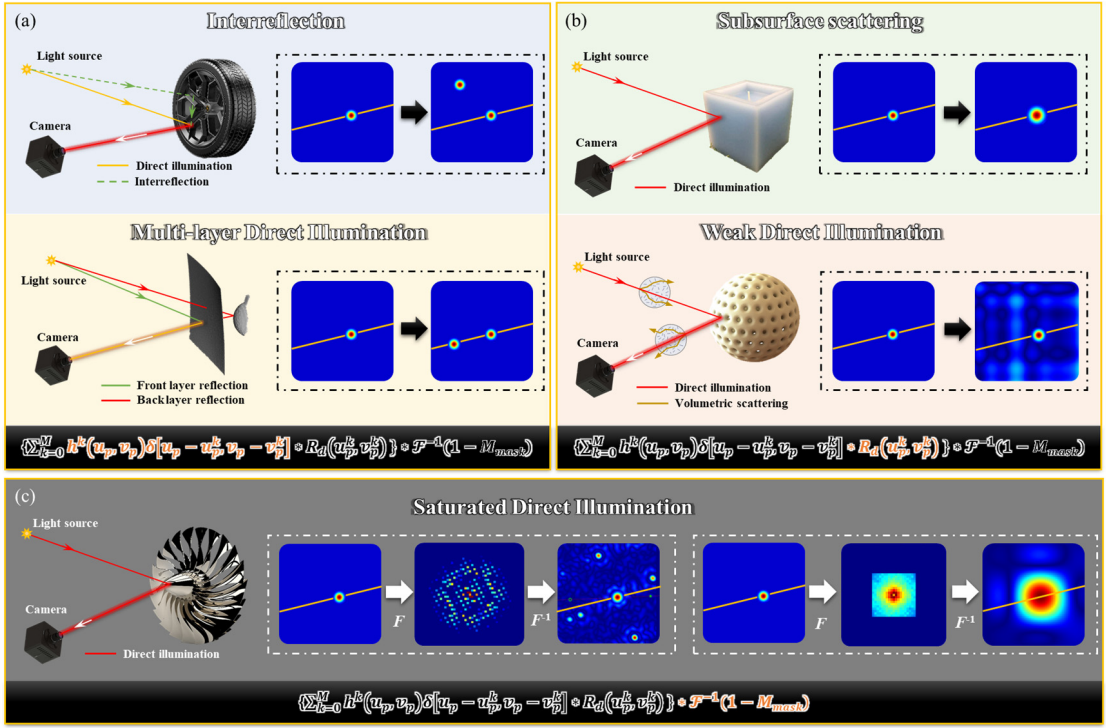

Figure 2. Imaging models under complex illumination. Impact of (a) inter-reflection and multipath illumination on the shifting factor, (b) subsurface scattering and thin scattering medium on the diffusion factor, and (c) overexposure on the sampling factor.

As shown in Fig. 2(a), shifting factor $\delta[u_p-u_p^k,v_p-v_p^k]$ produces multiple illumination components on LTC. But, interreflection illumination component falls out of the epipolar line because it breaks triangulation rule, while all multi-layer illumination components lie on the epipolar line because light transmission path does not change. Therefore, the imaging model of PSI can eliminate the influence of interreflection by rejecting indirect illumination components (fall out of epipolar line) and enables multi-layer surface measurement by respectively reconstructing all existing direct illumination components (lie on the epipolar line).

As shown in Fig. 2(b), the diffusion factor $R_d^k\left(u_p^k,v_p^k\right)$ can be expressed as:

$$R_d^k(u_p^k,v_p^k) = S(u_p^k,v_p^k)\frac{e^{-\frac{r}{d}} + e^{-\frac{r}{3d}}}{8\pi dr}, \qquad (7)$$

where $S(u_p,v_p)$ denotes the surface albedo; $r$ is the distance between incident and emergent points; $d$

depicts the mean free path of the scattering medium. Obviously, $R_d^k(u_p^k,v_p^k)$ is a decreasing and symmetrical function. Although LTC becomes diffused when suffering subsurface and volumetric scattering, the position of the center remains the same. Therefore, imaging model of PSI makes it insensitive to scattering properties of different surfaces and transmission properties of propagation medium.

As shown in Fig. 2(c), the sampling factor $M_{mask}$ illustrates the integrity of damaged spectrum coefficients. In cases where the surface exhibits excessively high reflectivity, the intensity of the acquired fringe patterns remains constant, reaching the maximum value detectable by the sensor, which is a phenomenon of overexposure. It causes the value of the spectrum coefficients calculated by Eq. (4) deviate from the theoretical value or even to zero. After the error shifts to the spatial domain, it results in the displacement and diffusion of the correct components. However, through multiple shifts and averaging, the correct direct component remains dominant and maximize the probability of correct localization. Therefore, the PSI imaging model makes it insensitive to high exposure surfaces.

The imaging model characterizes the variations in the "point-to-plane" LTC under complex reflection and transmission conditions through shifting, diffusion, and sampling factors. It demonstrates the general applicability of PSI, indicating its adaptability to a wider range of applications and laying the theoretical model for further optimization on demanding challenging tasks.

To validate the effectiveness of the imaging model and demonstrate the high adaptability of the PSI method, experiments were conducted under various complex reflection and transmission conditions, as shown in Fig. 3, including inter-reflections (highlighted by the red, orange, and blue boxes) and saturated direct illumination (marked by the green and yellow boxes) in a metallic scene, subsurface scattering in a jade qilin, direct light aliasing in a two-layer scene, and light scattering through thin smoke.

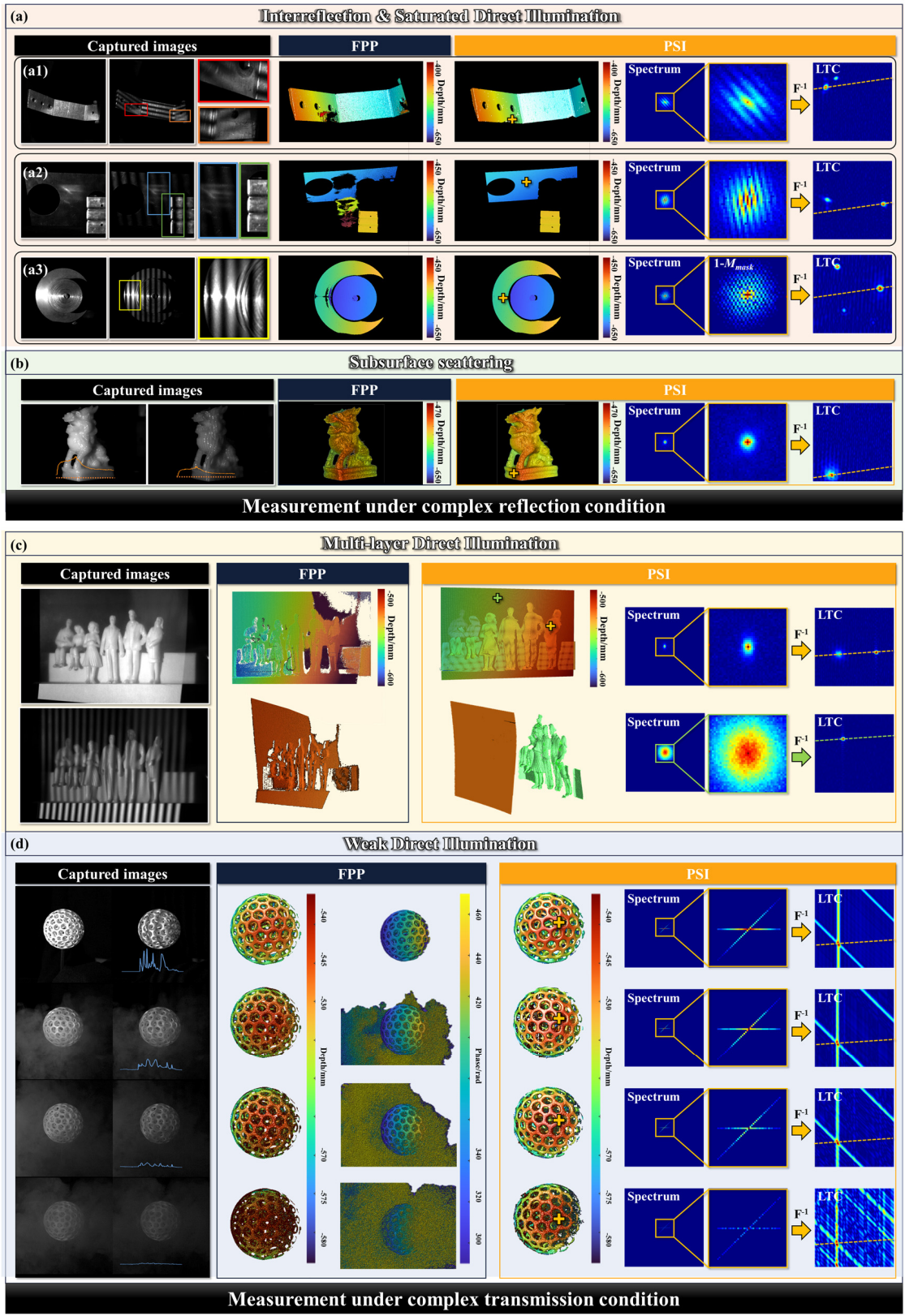

Figure 3. Measurement under complex reflection and transmission conditions using FPP and PSI methods. Captured images, 3D reconstruction results and LTC analysis for special pixels of scenes with (a) mutual reflections (a1 and a2) and overexposure (a3), (b) subsurface scattering, (c) a two-layer scene through a semitransparent surface, and (d) scenes through thin volumetric scattering media.

In these cases, the traditional FPP method fails because these conditions lead to phase damage or light aliasing issues, causing severe phase errors. In contrast, the PSI method demonstrates excellent consistency and robustness. Special pixels (highlighted by the plus signs) from the measurement scenes and their spectrum and LTC are presented for analysis. The corresponding epipolar lines on the LTC plane are represented by dashed lines. In the case of inter-reflection, the LTC consists of two illumination components: the direct illumination along the epipolar line and the global illumination resulting from inter-reflection off the epipolar line, as shown in Figs. 3(a1) and 3(a2). In overexposure conditions, the spectrum sampled as an irregular pattern, with the LTC showing multiple extreme points, where only the direct illumination component lies on the epipolar line, as shown in Fig. 3(a3). Thus, the direct illumination components can be accurately located. For the jade qilin with significant subsurface scattering, the profile lines of the white field and high-frequency fringe images are shown in Fig. 3(b). The fringe contrast is significantly reduced, becoming nearly invisible at high frequencies. The direct illumination component on the LTC exhibits scattering, but the central position remains nearly unchanged, ensuring successful positioning. In the two-layer scene with multiple aliasing direct components, as shown in Fig. 3(c), the LTC for the double-layer region (marked by the yellow plus sign) shows two direct illumination components on the epipolar line, while for the single-layer region (highlighted by the green plus sign) shows only one direct illumination component on the epipolar line, consistent with the derived imaging model. Hence, PSI method can separate and utilize the two layers of direct illumination to reconstruct the dual-layer scene. For the scene of a hollow sphere obstructed by spreading thin smoke, as shown in Fig. 3(d), an advanced MS-PSI method [45] with efficient projection was used for the experiment to avoid motion blur. Both FPP and PSI initially provide satisfactory results without smoke, but PSI shows higher accuracy. However, as the SNR decreases, the FPP results are significantly affected, as the phase are easily influenced by the presence of the smoke. While the PSI results are almost unaffected. Although the global components of the light transport function increase, the direct illumination component can still be accurately located.

In summary, the experimental results are consistent with the theoretical model, confirming its validity. And the proposed imaging model reveals the core principles and theoretically explains the intrinsic mechanisms for the superior performance of PSI under complex illumination conditions, providing a solid foundation for further optimization and development.

## 2.3 Robust and consistent measurement on low-SNR scenes with location-based noise model

Traditional PSI needs traverse the spectrum for reconstruction of LTC, so a large number of fringes are projected, as shown in Fig. 4(a). The LTC $h(u_p,v_p;u_c,v_c)$ is an ideal Dirac delta function $\delta$, representing the position of the direct illumination component, as shown in Fig. 4(b). However, due to the sparsity of the spectrum, only low-frequency spectral information needs to be collected. The LTC is filtered by a low-pass filter, degenerating to the generalized LTC $t(u_p,v_p;u_c,v_c)$, as shown in Fig. 4(d). Here, a rectangular low-pass filter form is considered, the red region indicates permitted transmission, as shown in Fig. 4(c). Noise may introduce erroneous global illumination components, so the goal is to locate the direct illumination component. The main lobe peak of $t(u_p,v_p;u_c,v_c)$ represents the correct direct illumination component, while the adjacent side lobe peak is typically the component closest to the main lobe, which is the most likely to interfere with the localization. Based on this, only the energy of the main lobe peak $t_{max}$ and the adjacent side lobe peak $t_{secmax}$ of $t(u_p,v_p;u_c,v_c)$ need to be considered, can be expressed as follows, and the detailed theoretical formula derivation is presented in Section 4.2:

$$t_{\max} = \lim_{u_p \to 0, v_p \to 0} g(u_p,v_p;u_c,v_c) = \frac{K'L'}{KL}, \tag{8}$$

$$t_{\sec\max} = t(\frac{5K'}{2K'},0;u_c,v_c) = \frac{1}{\sin(\frac{5\pi}{2K'})} \cdot \frac{L'}{KL}, \tag{9}$$

where $g(u_p,v_p;u_c,v_c)$ represents the spatial response of the low-pass filter, $K'$ and $L'$ represent the length in the horizontal and vertical directions respectively, $K$ and $L$ represent the total spectrum resolution.

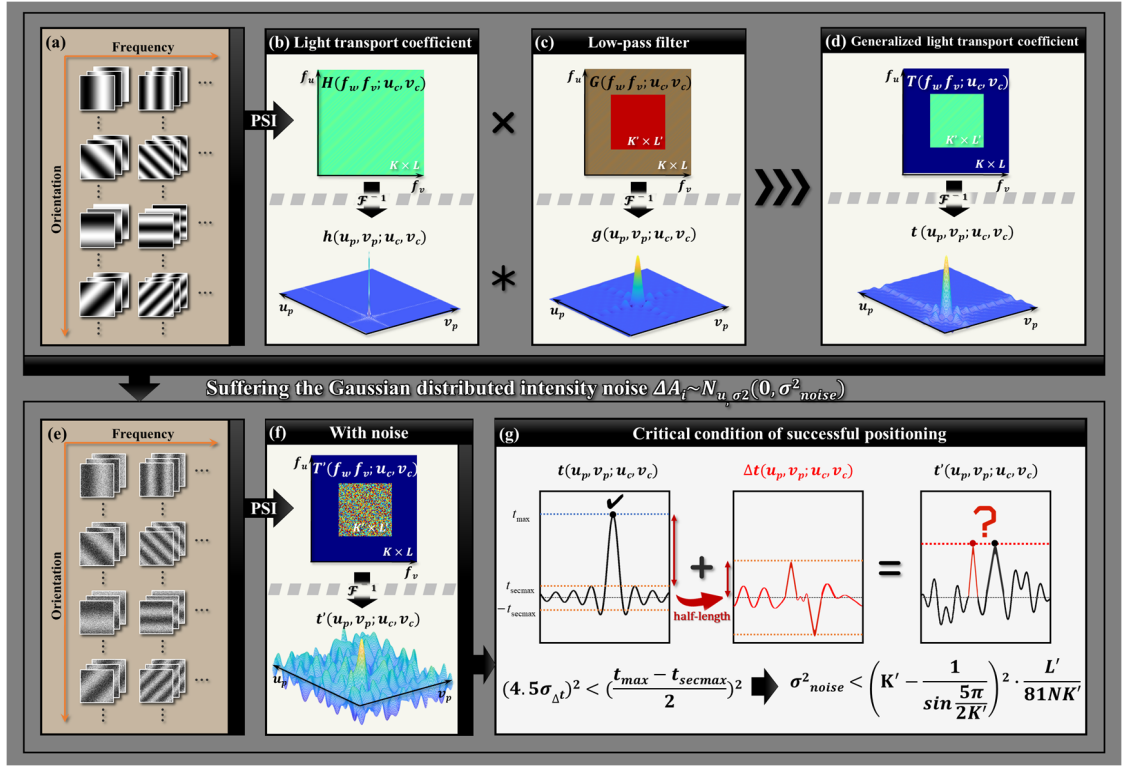

Figure 4. Illustration of the noise model. (a) Fringes without noise interference. (b) Ideal LTC for full-spectrum sampling. (c) Ideal rectangular low-pass filter. (d) Generalized LTC after low-pass sampling. (e) Fringes affected by noise. (f) Generalized LTC affected by noise. (g) Critical condition of successful positioning.

In structured-light measurement systems, most noise follows a Gaussian distribution, including noise from uneven illumination and thermal fluctuations of the camera's charge carriers [49]. And this work adopts the noise model proposed by Li et al [50], assuming the captured image experiences Gaussian noise $\Delta A_i$, where the mean is $\mu=0$ and the variance is $\sigma^2_{noise}$, as shown in Figs. 4(e)-4(g). Considering the intensity noise, the noise-affected generalized LTC $t'(u_p,v_p;u_c,v_c)$ is the original generalized LTC $t(u_p,v_p;u_c,v_c)$ with an added noise term $\Delta t(u_p,v_p;u_c,v_c)$. And the variance of $t(u_p,v_p;u_c,v_c)$ can be expressed in Eq. (10), and the explicit theoretical analysis is given Section 4.3.

$$\sigma_{\Delta t}^2 = N\sigma^2_{noise} \cdot \frac{K'L'}{(KL)^2}. \tag{10}$$

As shown in Fig. 4(g), once the maximum value of the superposition of $\Delta t(u_p,v_p;u_c,v_c)$ and $t(u_p,v_p;u_c,v_c)$ overwhelms the maximum value of $t(u_p,v_p;u_c,v_c)$, the positioning of the direct component will be wrong. Assume that a negative and positive noise appear at the position corresponding to the maximum value $t_{max}$ and the second maximum value $t_{secmax}$, respectively, and the noise magnitudes are both $(t_{max}-t_{secmax})/2$, the direct component is just drowned by noise. Since $\Delta t(u_p,v_p;u_c,v_c)$ follows a Gaussian distribution, the

4.5 sigma rule of thumb is used to define successful positioning [51]. In this case, the probability of correct positioning surpasses 99.9993%, ensuring near certainty. Thus, the critical condition of successful positioning can be expressed as:

$$(4.5\sigma_{\Delta t})^2 < \left(\frac{t_{\max} - t_{\sec\max}}{2}\right)^2. \tag{11}$$

Substituting Eqs. (8)-(10) into Eq. (11), this condition of successful positioning can be rewritten as:

$$\sigma^2_{noise} < \left(K' - \frac{1}{\sin\frac{5\pi}{2K'}}\right)^2 \cdot \frac{L'}{81NK'}. \tag{12}$$

It can be seen from Eq. (12), the maximum tolerable noise variance is only proportional to the product of the 2D spectrum sampling resolution.

To demonstrate the effectiveness of the noise model, simulations were performed for conditions with different sampling resolutions and varying levels of noise. The total resolution is selected as 201×201 pixels, and the sampling resolution ranges from 5×5 pixels to 141×141 pixels. After traversing each pixel to locate the correct component, the positioning accuracy reaches 99.9% as the simulation tolerance, as shown by the red lines in Fig. 5(a). The theoretical value calculated by Eq. (12) and the comparison of theoretical and simulated noise limits as shown in Fig. 5(b). The difference stems from the strict condition in theory, where the largest noises appear exactly at the main lobe peak and the adjacent side lobe peak positions. In simulation, random noise of the same level appears at other positions still allows correct localization, enabling to tolerate higher noise levels. Hence, the critical condition of noise tolerance in theory is more severe than simulation.

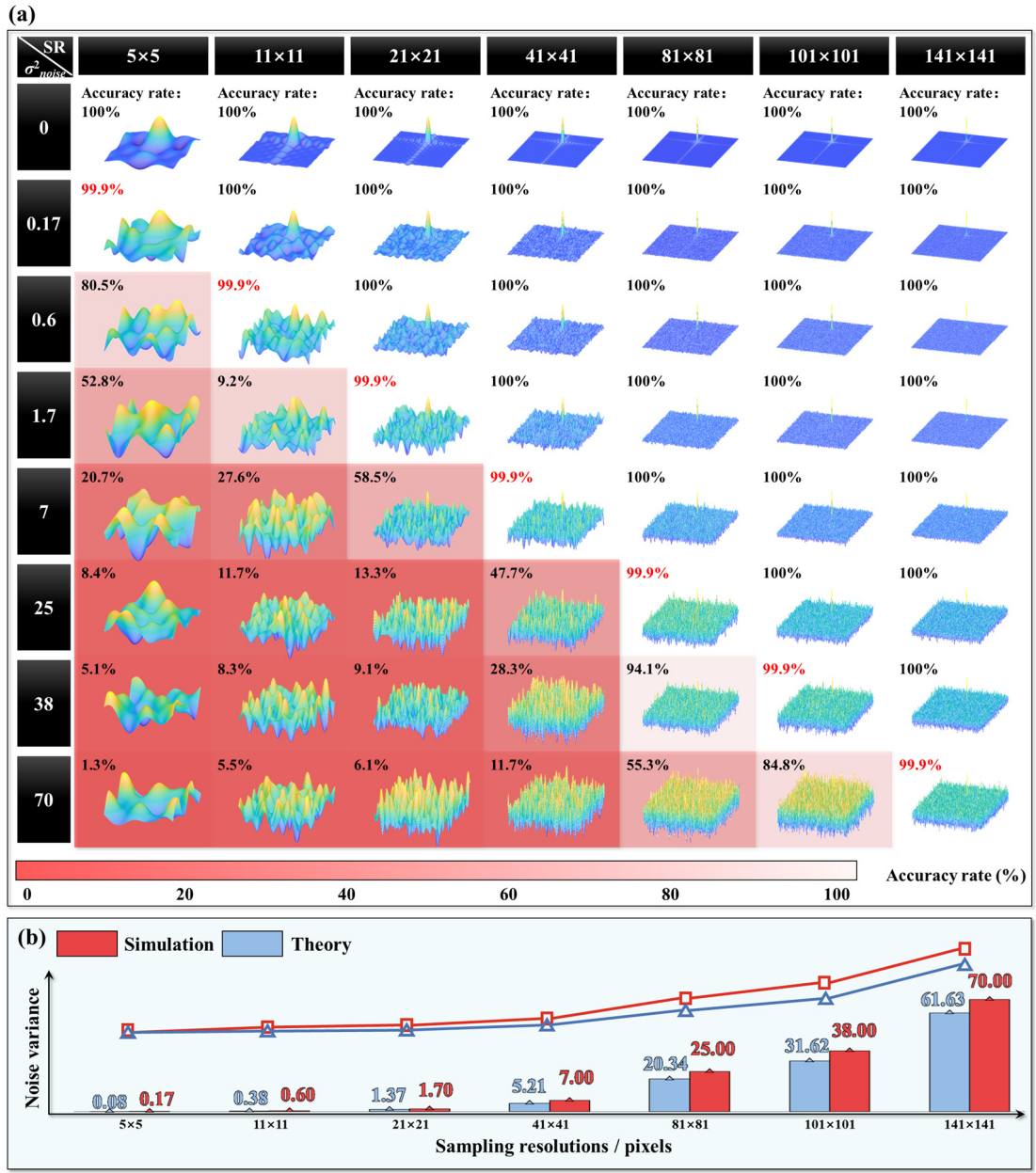

Figure 5. Simulations with different sampling resolutions and varying levels of noise. (a) LTC with different parameters. (b) Comparison of tolerance using the theoretical model and numerical simulation.

In order to verify the consistency between the theoretical model and actual experiments, accuracy evaluation experiments were conducted on two ceramic standard spheres with different reflectivity. The SNR is varied by adjusting the exposure time. The captured fringes with different SNR are shown in Fig. 6(a)-6(e), with the modulation decreasing from ~220 to ~7. The reconstruction results obtained using the traditional FPP method and the PSI method are shown respectively. The highest selected frequency is 45 fringe periods. To ensure fairness, all fringe images have been considered and used in the phase unwrapping process in FPP method. For the fringes with high SNR, both methods achieve good reconstruction, but PSI outperforms FPP with over twice higher precision (i. e. 52/23=2.26) as shown in Figs. 6(f) and 6(p). Approaching the critical SNR threshold where FPP begins to fail (around 0.06 dB),

FPP exhibits significant nonlinear errors in Fig 6(g), while PSI maintains high accuracy as shown in Fig. 6(q). As the SNR further decreases, FPP fails, while PSI still keeps a low error rate of 3.67%, much lower than 45.58% of FPP as shown in Figs. 6(o) and 6(y). It demonstrates PSI method indeed exhibits better noise resistance, with superior consistency and robustness.

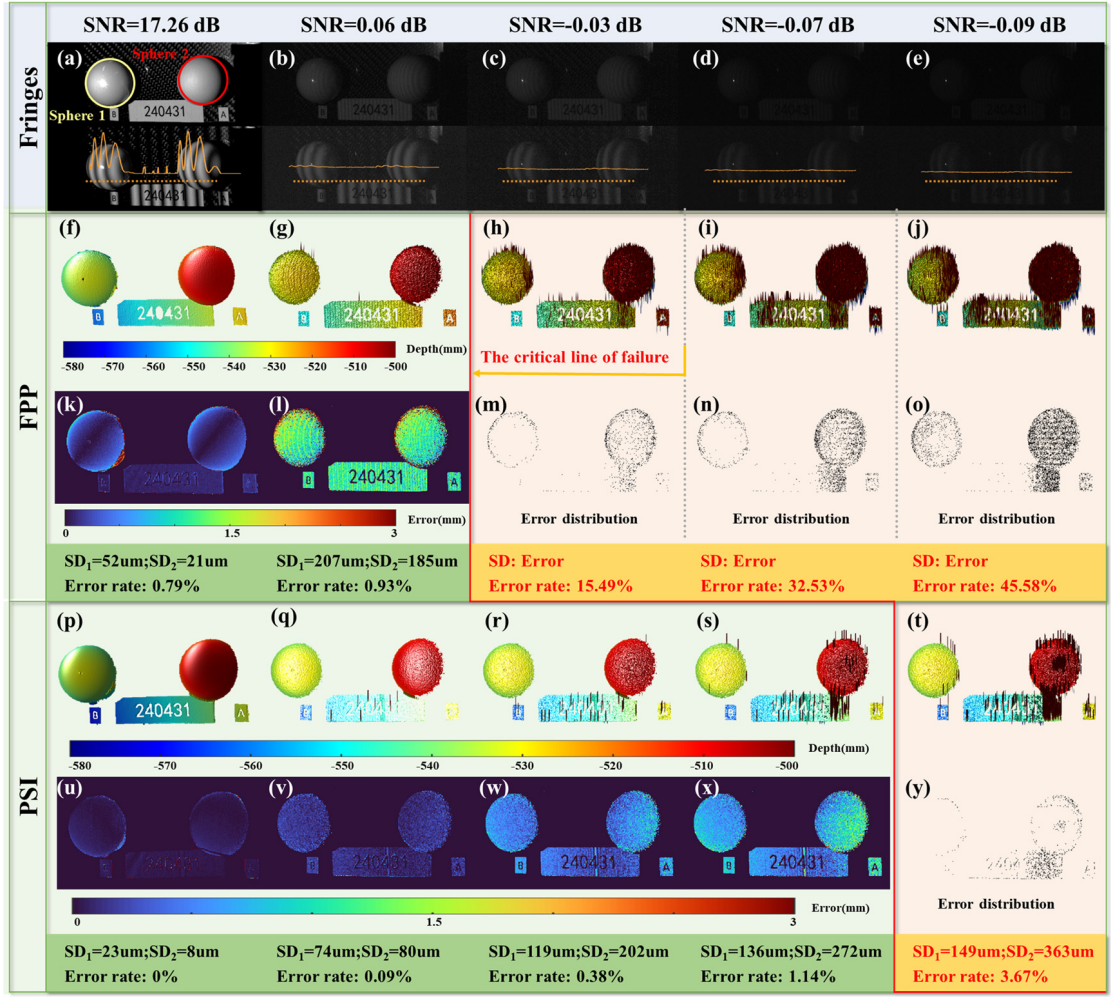

Figure 6. Accuracy evaluation experiments with different SNR. (a)-(e) Captured white field images and fringes with different SNR. (f)-(j) Reconstruction results using the FPP method. (k)-(l) Error map of FPP results. (m)-(o) Error distribution of FPP results. (p)-(t) Reconstruction results using PSI method. (u)-(x) Error map of PSI results. (y) Error distribution of PSI result.

Simulations and experiments validate the effectiveness of the noise model, revealing that PSI outperforms the traditional FPP method in terms of noise resistance, demonstrating significant potential under extreme low SNR conditions. The key difference lies in noise propagation. Traditional FPP entirely relies on phase quality and follows a point-to-point reconstruction, noise cannot be mitigated [49]. In contrast, PSI noise directly affects spectral coefficients and is distributed across the entire field by inverse Fourier transform, reducing its impact in the LTC and localization. Moreover, as the spectral sampling rate increases, the impact of noise in the spatial domain further diminishes. This underlying mechanism enables PSI to achieve improved accuracy under low SNR conditions.

## 3. Challenges and opportunities

PSI is still in its infancy. There remains much work to be done, as shown in Fig. 7. We will discuss some directions for future study of PSI in this section.

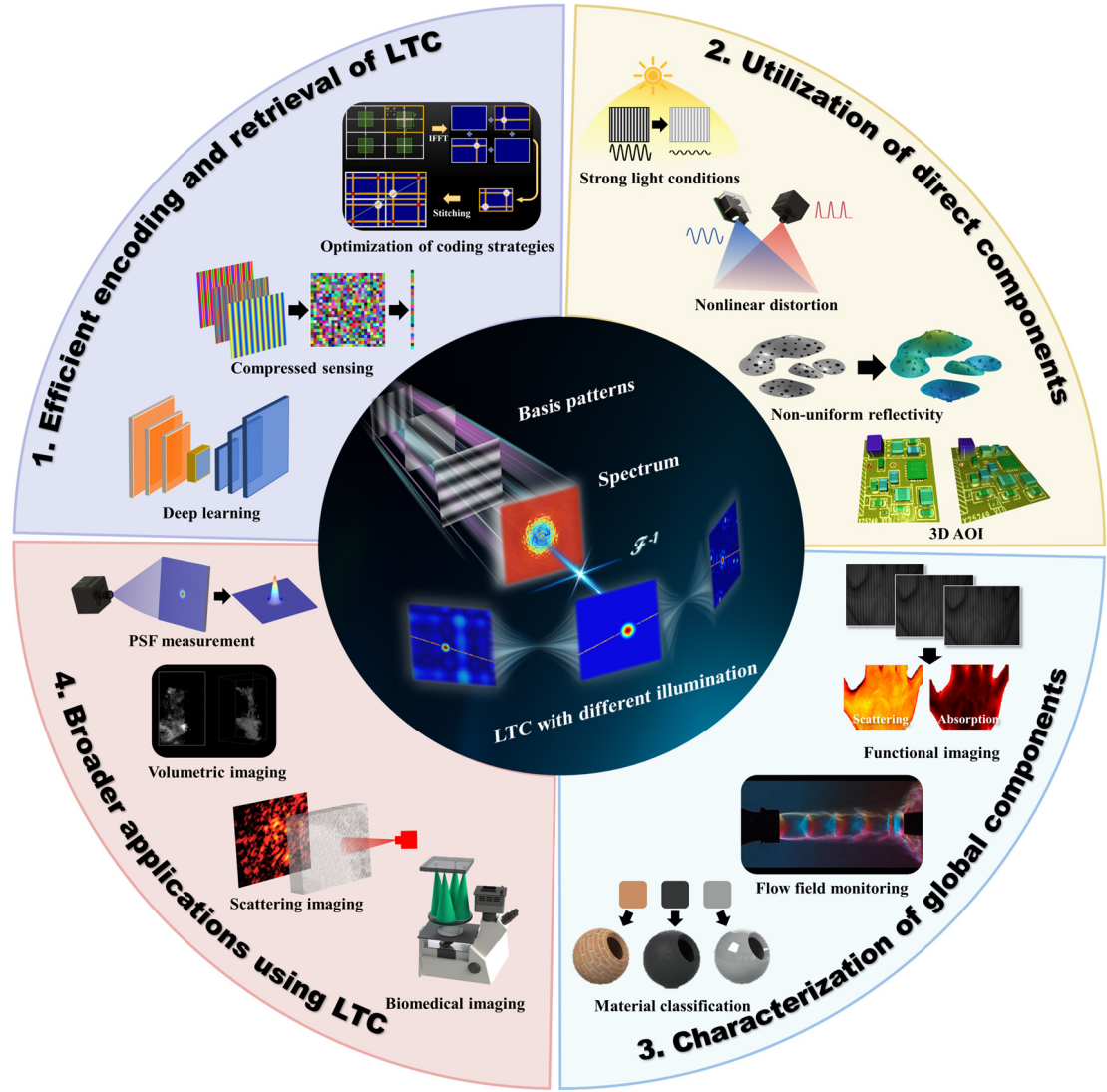

Figure 7. Challenges and opportunities of PSI.

### 3.1 Efficient encoding and retrieval of LTC

Compared to traditional structured light techniques, PSI shows significant superiority on measurement performance. However, its main drawbacks lie in two aspects:

- **Low coding efficiency of LTC:** it is necessary to traverse the entire spectrum and project plenty of patterns to encode LTC.

- **High computational cost of LTC:** the reconstruction process of LTC needs to perform single-pixel imaging algorithm pixel by pixel, resulting in high computational cost.

These limitations hinder PSI in high-speed and real-time imaging applications, which are critical in medical imaging, industrial inspection, and human-computer interaction. Thus, efficient algorithm design and processing is essential.

To improve coding efficiency, recent efforts have focused on projective parallel single-pixel imaging [52] and multi-scale parallel single-pixel imaging [45]. These methods reduce the number of projection patterns required for each reconstruction to as few as 15, greatly improving acquisition efficiency and supporting dynamic 3D imaging with PSI. Due to the point-to-plane nature of PSI, balancing acquisition efficiency and reconstruction accuracy remains challenging, and both cannot be optimized simultaneously. Infinite reduction of projection patterns is impractical, and further research is needed to refine projection strategies, including:

- **Sampling slice optimization based on scene frequency sensitivity:** different scenes exhibit varying sensitivities to the frequencies. Prioritizing the sampling of frequency components most relevant to the target scene helps maximize useful information under low sampling conditions. Additionally, the choice of sampling slice direction also impacts localization accuracy [46, 53].

- **Adaptive joint parameter optimization:** the number of projected patterns is determined by parameters such as the phase-shifting steps [54] of the projection patterns, the scale, and the sampling coefficients based on MS-PSI. Therefore, exploring the joint optimization of these parameters presents a potential approach to minimize the number of projected patterns.

- **Time multiplexed projection strategies:** by redesigning the phase shifts and pattern arrangement in the time sequence, and fully exploiting the correlation between adjacent images in high-speed recordings, the number of projections can be effectively reduced while enhancing resistance to motion blur [12].

To optimize the issue of high computational cost, improvements can be made in two main areas:

- **Algorithm optimization:** due to spectral sparsity, compressed sensing techniques [55, 56] can be employed to reduce time complexity, reducing reconstruction time. Deep learning [57, 58] also can enhance processing speed, such as automating feature extraction for rapid localization or accelerating computations through batch parallel processing.

- **Hardware acceleration:** utilizing graphics processing units (GPUs) for large-scale parallel computing, or using specialized hardware accelerators, can significantly improve computational efficiency.

## 3.2 Interference-resistant extraction and utilization of direct components

In addition to the complex illumination components analyzed in this study, several other important issues warrant further attention in PSI applications. In practice, more intricate scenarios may impose challenging conditions on the selection of the LTC. The presence of weak or multiple direct components greatly complicates accurate localization. These challenges can be broadly classified into the following three types of tasks.

- **Direct illumination discrimination among the shifting components:** due to the point-to-plane nature of the imaging model, PSI is inherently suited to address the issue of multiple direct illumination components overlapping at the same pixel. For instance, on surfaces with non-uniform reflectivity [59-61] may cause information aliasing; in step structures or regions with abrupt depth changes [53], boundary information may mix. In multi-layered scenes [62], direct components from different depths may overlap, including both specular [63] and diffuse reflection. Additionally, periodic signals in the spectrum can introduce shifting factor with higher-order harmonics on the LTC, such as ripple artifacts caused by nonlinear distortion [64] from the gamma effect. These violate the traditional point-to-point triangulation principle, whereas PSI offers a potential solution based on the position invariant of the direct component [45].

- **Direct illumination localization under low SNR :** when the scene is outside the device's depth of field [65] or the object's height varies significantly, defocus can cause fringe blurring, typically requiring additional compensation in traditional methods [66]. Similar to subsurface scattering, defocus effects manifest as uniform dispersion on the LTC, which can directly localize the direct components. Additionally, under strong light conditions (e.g., outdoor measurements), the reduced fringe contrast resembles the low contrast, can also be addressed using similar way.

- **Ambiguity elimination among the duplicating components:** unstable light sources alter fringe intensity and contrast, causing reconstruction errors in traditional FPP. However, this effect may be confined to a limited set of spectral coefficients in PSI, essentially forming an irregular error mask in frequency domain. And duplicating components emerge to disturb the location of direct component. Based on noise model analysis, the system generally exhibits high tolerance to spectral errors, suggesting that the PSI method is promising to deal with harsh environments with light source fluctuations [67].

## 3.3 Decoupling and characterization of global components

Currently, the direct component is most utilized, but point-to-plane imaging capability of PSI can also be leveraged to separate and use underexplored global components. The global components correspond to the off-epipolar illumination in the LTC. They are crucial for optical property measurement, particularly for the quantitative analysis of absorption and scattering in highly turbid media [68, 69]. Conventional spatial frequency domain imaging (SFDI) techniques typically rely on switching multi-wavelength coherent light source to independently extract the concentration of each chromophore, impeding fast measurement [70]. PSI system has potential to capture global components in an incoherent structured light system, offering a new insight for optical property analysis and material classification in fields such as online inspection, biomedical imaging and dynamic monitoring of fluid fields. However, accurate decoupling of global components and mapping optical properties remains the open challenges to be addressed:

- **Global components decoupling:** according to the imaging model in Section 2.2, global components may originate from inter-reflection effects, leading to additional variations in the shifting factor, or from subsurface scattering, causing speckle blurring and influencing the diffusion factor. Unlike direct illumination only requires positional information, the characterization of global illumination demands complete information, which represents the coefficients of absorption, scattering, and reflection. Thus, extracting complete global component information is crucial, especially when employing dimension-reduced projection strategies.

- **Global components characterization:** inter-reflection information reveals the deflected positions of real light paths, enabling realistic scene illumination simulation. Subsurface scattering coefficients and the distribution of defocused spots correspond to different biological tissues or materials, facilitating material identification and classification while also providing prior knowledge for parameter selection in efficient reconstruction. However, mapping global components to the coefficients of absorption, scattering, and reflection remains a critical issue to be solved.

### 3.4 Broader applications using LTC

Beyond direct and global component separation, the measured LTC holds significant potential for broader applications. Including, but not limited to, the following potential applications:

- **Characterization of optical system imaging quality:** the imaging process of optical systems is influenced by various factors, making it challenging to accurately characterize these effects. LTC measurement, similar to ray tracing, serves as a system characterization tool to describe the propagation and interaction of light, and can also be applied to non-standard projector calibration [71].

- **Points spread function (PSF) measurement:** LTC represents the PSF under spatial variations, accurately describing the light transmission process, enabling PSF measurement under large depth-of-field conditions [72].
- **Image deblurring and descattering:** LTC describes the light transmission process, making it crucial for deblurring and de-scattering imaging, such as underwater imaging [73], biomedical imaging [74] and volumetric imaging [75]. In microscopy, the combination of de-scattering with LTC can effectively extend the depth of field of measurements.

## 4. Conclusion

This paper establishes a comprehensive theoretical model for PSI for the first time, integrates both imaging and noise models. The imaging model mathematically represents the LTC under complex reflection and transmission conditions, revealing underlying mechanisms of PSI. It employs a computational imaging framework for point-to-plane reconstruction, moving beyond the traditional point-to-point phase analysis in structured light techniques. Meanwhile, the location-based noise model evaluates the impact of environmental noise on measurement accuracy, providing a systematic error analysis for PSI systems. Its point-to-plane imaging allows it to separate complex illumination and distribute noise, resulting in superior noise resistance and robustness over traditional structured light techniques. Comparative results demonstrate that simulations and experiments both align with the framework and conclusions of the proposed models, confirming the generality and robustness of PSI. Based on the comprehensive theoretical model, this work further explores potential challenges and opportunities PSI may address in the future and its prospective applications. These theoretical contributions lay a strong foundation for the continued advancement and practical application of PSI in increasingly complex 3D reconstruction tasks. The novel point-to-plane analysis perspective paves the way for endless opportunities for next-generation structured light 3D measurement technology.

## 5. Materials and Methods

### 5.1 Experimental setup details

The experiment utilizes a DLP projector (DLP VisionFly 6500) with a 1920 × 1080 resolution and a high-speed camera (Photron FASTCAM Mini AX200). The camera is equipped with a Nikon AF-S 28-300 mm f/3.5-5.6 G ED VR zoom lens, set to a 35 mm focal length with an F8 aperture. The camera was synchronized with the projector using a trigger signal, and its resolution was set to 1024 × 672 pixels for all experiments.

## 5.2 Derivation of the maximum and second maximum values of the generalized LTC

Due to the sparsity of the spectrum, only low-frequency spectrum information can be collected, Eq. (4) can be rewritten as:

$$T(f_u, f_v; u_c, v_c) = H(f_u, f_v; u_c, v_c) \cdot G(f_u, f_v; u_c, v_c), \quad (13)$$

where $T(f_u, f_v; u_c, v_c)$ is defined as the generalized spectral coefficients, and $G(f_u, f_v; u_c, v_c)$ represents a rectangular low-pass filter, which can be written as:

$$G(f_u, f_v; u_c, v_c) = \begin{cases} 1, & \text{if } |f_u| \leq \frac{K'}{2} \text{ and } |f_v| \leq \frac{L'}{2}, \\ 0, & \text{otherwise.} \end{cases} \quad (14)$$

The generalized LTC can be written as:

$$t(u_p, v_p; u_c, v_c) = h(u_p, v_p; u_c, v_c) * g(u_p, v_p; u_c, v_c), \quad (15)$$

where $g(u_p, v_p; u_c, v_c)$ represents the spatial response of the low-pass filter, can be written as:

$$g(u_p, v_p; u_c, v_c) = \frac{1}{KL} \sum_{f_u=-K'/2}^{K'/2} \sum_{f_v=-L'/2}^{L'/2} G(f_u, f_v; u_c; v_c) e^{-j2\pi \frac{u_p f_u}{K}} e^{-j2\pi \frac{v_p f_v}{L}}$$

$$= \frac{1}{KL} \cdot \frac{\sin(\pi \frac{u_p K'}{K})}{\sin(\pi \frac{u_p}{K})} \cdot \frac{\sin(\pi \frac{v_p L'}{L})}{\sin(\pi \frac{v_p}{L})}. \quad (16)$$

It can be considered as the product of two one-dimensional functions. The zero points of a one-dimensional can be represented as:

$$u_p = \frac{mK}{K'}, m \in Z. \quad (17)$$

The zero-order zero point corresponds to the main lobe peak $g_{max}$, the middle between the second-order and third-order zero-crossing points approximately corresponds to the adjacent side lobe peak $g_{secmax}$. And it can be expressed as:

$$g_{max} = \lim_{u_p=0, v_p=0} g(u_p, v_p; u_c, v_c) = \frac{K'L'}{KL}, \quad (18)$$

$$g_{\text{sec max}} = g(\frac{5K'}{2K},0;u_c,v_c) = \frac{1}{\sin(\frac{5\pi}{2K'})} \cdot \frac{L'}{KL}. \tag{19}$$

According to Eq. (15), the main lobe peak $t_{max}$ and the adjacent side lobe peak $t_{secmax}$ of $t(u_p,v_p;u_c,v_c)$ can be expressed as:

$$t_{\max} = g_{\max} = \frac{K'L'}{KL}, \tag{20}$$

$$t_{\text{sec max}} = g_{\text{sec max}} = \frac{1}{\sin(\frac{5\pi}{2K'})} \cdot \frac{L'}{KL}. \tag{21}$$

### 5.3 Derivation of variance of the transmitted noise

Considering the Gaussian distributed intensity noise $\Delta A_i$ with a mean value $\mu=0$ and a variance $\sigma^2_{noise}$, as shown in Figs. 4(e)-4(f), the patterns in Eq. (3) can be rewritten as:

$$I'_i(f_u,f_v;u_c,v_c) = A(u_c,v_c) + \sum_{v_p=0}^{L'-1}\sum_{u_p=0}^{K'-1} t(u_p,v_p;u_c,v_c)P_i(u_p,v_p;f_u,f_v) + \Delta A_i(f_u,f_v;u_c,v_c). \tag{22}$$

And the spectrum affected by noise and after low-pass filtering can be rewritten as:

$$T'(f_u,f_v;u_c,v_c) = [\sum_{i=1}^{N} I'_i(f_u,f_v;u_c,v_c)\sin\delta_i] + j[\sum_{i=1}^{N} I'_i(f_u,f_v;u_c,v_c)\cos\delta_i]$$
$$= T(f_u,f_v;u_c,v_c) + \Delta T(f_u,f_v;u_c,v_c), \tag{23}$$

where $\Delta T(f_u,f_v;u_c,v_c)$ represents the noise term of the spectrum, and it is written as:

$$\Delta T(f_u,f_v;u_c,v_c) = [\sum_{i=1}^{N} \Delta A_i(f_u,f_v;u_c,v_c)\sin\delta_i] + j[\sum_{i=1}^{N} \Delta A_i(f_u,f_v;u_c,v_c)\cos\delta_i]$$
$$= a(f_u,f_v;u_c,v_c) + jb(f_u,f_v;u_c,v_c), \tag{24}$$

where $a(f_u,f_v;u_c,v_c)$ and $b(f_u,f_v;u_c,v_c)$ represent the real and imaginary parts of the noise term respectively.

After inverse Fourier transform, the generalized LTC affected by noise can be written as:

$$t'(u_p,v_p;u_c,v_c) = F^{-1}[T'(f_u,f_v;u_c,v_c)] = \frac{2 \cdot \{t(u_p,v_p;u_c,v_c) + F^{-1}[\Delta T(f_u,f_v;u_c,v_c)]\}}{Nb}, \tag{25}$$

where the response of the noise in the spatial domain can be written as:

$$\Delta t(u_p,v_p;u_c,v_c) = \frac{1}{KL} \sum_{f_u=-K'/2}^{K'/2} \sum_{f_v=-L'/2}^{L'/2} \Delta T(f_u,f_v;u_c;v_c) e^{-j2\pi\frac{u_p f_u}{K}} e^{-j2\pi\frac{v_p f_v}{L}}$$
$$= \frac{1}{KL} \sum_{f_u=-K'/2}^{K'/2} \sum_{f_v=-L'/2}^{L'/2} [a(f_u,f_v;u_c;v_c) + jb(f_u,f_v;u_c;v_c)]e^{-j\theta}, \tag{26}$$

where $\theta = 2\pi(\frac{u_p f_u}{K} + \frac{v_p f_v}{L})$. According to the conjugate symmetry property of the spectrum, one of the conjugate pairs can be extracted and written as:

$$C_i = [a(f_u, f_v; u_c; v_c) + jb(f_u, f_v; u_c; v_c)]e^{-j\theta} + [a(-f_u, -f_v; u_c; v_c) + jb(-f_u, -f_v; u_c; v_c)]e^{j\theta}$$
$$= 2[a(f_u, f_v; u_c; v_c) \cdot \cos\theta - b(f_u, f_v; u_c; v_c) \cdot \sin\theta] \quad (27)$$
$$= 2[\sum_{i=1}^{N} \Delta A_i \sin(\delta_i - \theta)].$$

Since $\Delta A_i$ follows a Gaussian distribution with variance $\sigma^2_{noise}$, $C_i$ also follows a Gaussian distribution based on the properties of Gaussian error propagation [76]. And the variance $\sigma_{Ci}^2$ of $C_i$ can be described as:

$$\sigma_{C_i}^2 = 4 \cdot \sigma^2 \sum_{i=1}^{N} \sin^2(\delta_i - \theta) = 4 \times \frac{N\sigma^2}{2}. \quad (28)$$

In addition, the real term of the spectrum can be written in the form of Eq. (28):

$$R_i = a(f_u, f_v; u_c; v_c)e^{-j\theta} = a(f_u, f_v; u_c; v_c) \cdot \cos\theta - b(f_u, f_v; u_c; v_c) \cdot \sin\theta = \frac{1}{2}C_i. \quad (29)$$

Thus, the variance $\sigma_{Ri}^2$ of $R_i$ can be described as:

$$\sigma_{R_i}^2 = \frac{N\sigma^2}{2}. \quad (30)$$

Therefore, the variance $\sigma_n^2$ of the optical transport coefficients fluctuation $n(u_p,v_p;u_c,v_c)$ caused by noise is related to the number of conjugate pairs and real terms of the sampled spectrum. When $K'$ and $L'$ is much larger than 4, $\sigma_n^2$ can be approximately expressed as:

$$\sigma_{\Delta t}^2 = N\sigma^2_{noise} \cdot \frac{K'L'}{(KL)^2}. \quad (31)$$

## Acknowledgments

Authors thank Prof. Hongzhi Jiang from Beihang University and Prof. Chao Zuo from Nanjing University of Science and Technology for their helpful discussion on theoretical models of PSI in September 2024 at Computational Imaging Conference (CITA2024).

## Author contributions

Z. J. Wu and Q. C. Zhang supervised the study. Z. J. Wu, F. F. Chen and Y. N. Shen conceptualized the research. Z. J. Wu and F. F. Chen conceived, designed and performed the experiments. F. F. Chen, Y. N. Shen, C. M. Liu, Z. S. Chen and T. Xi developed the theoretical derivation. F. F. Chen and Z. D. Chen developed the numerical methods. All authors participated in the data analysis and contributed to the writing of the manuscript.

## Funding

This research was supported by the National Natural Science Foundation of China (62205226), the National Postdoctoral Program for Innovative Talents of China (BX2021199), the General Financial Grant from the China Postdoctoral Science Foundation (2022M722290), and the Key Science and Technology Research and Development Program of Jiangxi Province (20224AAC01011).

## Data availability

Data available on request from the authors.

## Conflict of interest

The authors declare no competing interests.